\documentclass{pasj00}
\usepackage{graphicx}
\draft

\begin{document}
\SetRunningHead{H.Kamaya}{SDSS J1257+3419}
\Received{2007/06/04}
\Accepted{2007/08/22}

\title{The faint stellar object SDSS J1257+3419 is a dark matter dominated system}

\author{Hideyuki \textsc{kamaya} 
}

\affil{Department of Earth and Ocean Sciences, School of Applied Sciences, 
National Defense Academy of Japan, \\
Yokosuka 239-8686, Japan}
\email{kamaya@nda.ac.jp}


%


\maketitle

\begin{abstract}

SDSS J1257+3419 has been reported as either a faint and small dwarf galaxy or a faint and widely extended globular cluster. In this Letter, the author insists this stellar system is a dwarf spheroidal (dSph). Adopting an observational relation between binding energy and mass of old stellar systems, we derive a new relation between mass and size of dSphs by assuming that they are dark matter dominated and virialized objects. 
Letting half-light radius represent size of SDSS J1257+3419, we find that its mass is $\sim 7\times 10^6$ solar mass. 
This indicates mass-to-light ratio ($M/L$) of SDSS J1257+3419 is about 1000 in the solar unit. This large $M/L$ is expected from a Mateo plot of dSphs. 
Thus, we confirm SDSS J1257+3419 is a dSph.

\end{abstract}

\section{Introduction}

There are many dwarf spheroidals (dSphs) around the Milky Way.
They are known to be low-luminosity,
low-surface-brightness dwarf elliptical galaxies
(e.g. Irwin \& Hatzidimitriou 1995).
Their observational features are the following:
absolute visual magnitude is fainter than $-14$, and 
shape is near spherical. 
The classical studies often thought that dSphs were merely
large, low-density globular clusters. Recent studies have shown that
dSphs have a more complex stellar population than what is found in
globular clusters (e.g. Unavane, Wyse, \& Gilmore 1996).
Furthermore, although globular clusters have
distinct nucleus, dSphs do not have. 
It is established that dSphs are very different from globular clusters.

Generally, in dSphs, star
formation over extended periods is expected. 
This is true although some of them show no sign of
current or recent star formation and have no detectable interstellar
matter (e.g. van den Bergh 1999). 
There are two main stellar population : 
old metal-poor stars which are similar to those of
Galactic globular clusters, 
and intermediate-age stars, whose ages range
from one to 10 billion years. The first population indicates
the starburst 
(e.g. Gallart et al. 1999) which triggers strong gas outflow (e.g. galactic wind).

According to Saito (1979a, b)
and the recent paper by de Rijcke et al. (2005), in dSphs
around the Milky Way, the galactic wind
occurs (Dekel and Silk 1986). This is concluded as a fact 
that their stellar mass density is
relatively low although their mass is comparable to that of globular
clusters. This also means that there is the ejected interstellar medium from
dSphs in the Milky Way halo (e.g. Mayer et al. 2006). 
That is, the halo medium is polluted by metal supplied by
 the galactic winds of dSphs. 
Then, it is meaningful to detect dSphs around the Milky
Way. Especially, we need to know the number of very small
(e.g. Read, Pontzen, \& Viel 2006).

There are a lot of dSphs discovered recently (e.g. Gilmore et al. 2007). 
Especially, the discoveries of very faint dSphs are important to study the general property of dSphs. The representative faint dSphs are Ursa Major II (Zucket et al. 2006) and Coma Berenices (Belokurov et al. 2007).
A recent analysis (Sakamoto \& Hasegawa 2006; Belokurov et al. 2007) has also shown there is a very small object at 150 kpc from the Sun: SDSS J1257+3419 which is named as Canes Venatici II. 
This system can be either a
faint and small dwarf galaxy like dSphs or a faint and widely extended globular
cluster according to the stellar properties of the system. 
Although Kamaya (2007) suggests SDSS J1257+3419 is one of dSphs 
 because of its sparse stellar density,  
 there was no distinct answer since its mass-to-light ratio is not obtained well.
 
 The difficulty originates from the fact that SDSS J1257+3419 is about three times further than Ursa Major II and Coma Berenices. By the way, the fainter dSphs are, the larger mass-to-light ratio ($M/L$) is as shown in figure 5 of Gilmore et al. (2007), which is called a Mateo plot (Mateo et al. 1993). For example, $M/L$ of Bootes dSph, which absolute magnitude is about -6 in V-band, is about 600 in the solar unit (Gilmore et al. 2007). According to the Mateo plot, if we reveal that $M/L$ of SDSS J1257+3419, which absolute magnitude is -4.8 in V-band, is about 1000, we can insist it is a dSph.
Thus, in this Letter, we try to estimate $M/L$ of SDSS J1257+3419.


\section{Formulation}

Fist of all, defining a proto-dSph, we adopt the following assumptions. 
(1) dSphs form like globular clusters and/or elliptical galaxies at their birth epoch. 
This means proto-dSphs obey the relation of Saito (1979a): 
$$E_{\rm b} = 3.4 \times 10^{-6} (M_0)^{1.45}  {\rm erg} . 
\eqno(1)$$ 
when they form, where $E_{\rm b}$ is binding energy and $M_0$ is total mass of the system in cgs units. 
Just before the galactic wind era, proto-dSphs are still gas rich systems. 
(2) After that, proto-dSphs deviate from this relation owing to the galactic wind.
(3) Finally, present dSphs are re-virialized. 

Total energy of a system is half of binding energy, 
$E_{\rm b}/2$, when the system is virialized. 
The fractional energy of the dark matter component, then, 
$$E_{\rm DM} = \frac{M_{\rm DM} }{ M_{\rm 0} } \frac{E_{\rm b}}{2} 
\eqno(2).$$
{\bf We define $E_1$ as the total energy of a dSph after the galactic wind era.} 
Because almost gas is blown away 
and stellar mass is expected to be much smaller than $M_{\rm DM}$,
$E_1$ is nearly equal to $E_{\rm DM}$.
{\bf We also assume $M_0 \sim M_{\rm DM}$ since dSphs always have large $M/L$} (e.g. Hirashita, Takeuchi, \& Tamura 1998). 
As long as the system is re-virialized, $E_1 = -GM_{\rm DM}^2/4 R_1$
where $R_1$ is radius of a present dSph. 
In cgs units, then, 
we derive a relation between $R_1$ and $M_{\rm DM}$:  
$$ R_1 = 0.98 \times 10^{-2} \times {M_{\rm DM}}^{0.55} . 
\eqno(3)
$$

\section{Results and Discussion}

We shall check the above relation between $M_{\rm DM}$ and radius of $R_1$.
The sample dSphs are listed in table 1. 
For convenience, some basic physical quantities are summarized.
They are chosen because half-light radius, $r_{1/2}$, and
$M/L$ are determined well. 
Then, well-known dSphs Ursa Major
and Sagittarius are omitted from this list. Obviously, SDSSJ1257 denotes
SDSS J1257+3419.
Here, we call the other nine sample classical dSph. 

\begin{table}[t]
  \begin{tabular}{|c|c|c|c|c|c|}

\hline

  dSph &  distance (kpc) & Mv & half light radius (kpc) & M/L & $R_1$ (kpc) \\

\hline

SDSSJ1257  & 150   &    -4.8 &  0.038 & --   & -- \\

Bootes         &  60    &   -5.7  &  0.22 & 555 & 0.058 \\
UrsaMinor      & 66     &   -8.9   &  0.15 & 95 & 0.094 \\
Sculptor       &  79    &  -11.1   &  0.094 & 11 & 0.090 \\
Draco          &  82    &   -8.8  &  0.12 & 243 & 0.113 \\
Sextans        &  86    &   -9.5  &  0.29 & 107 & 0.116 \\
Carina         & 101    &   -9.3  &  0.14 & 59 & 0.074 \\
Fornax         & 138    &  -13.2   &  0.34 & 7 & 0.172 \\
LeoII          & 205    &   -9.6   &  0.12 & 23 & 0.200 \\
LeoI           & 250    &  -11.9   &  0.12 & 1.0 & 0.031 \\

\hline
  \end{tabular}
\caption{dSph sample: 
Data are taken from Irwin \& Hatzidimitriou (1995), 
except $r_{1/2}$ of Bootes (Zucker et al. 2006; Belokurov et al. 2006), 
and M/L of Bootes (Mu\~{n}oz et al. 2006; Belokurov et al.; Gilmore et al. 2007).
M/L of Bootes is estimated by using absolute magnitude of Belokurov et al. and dynamical mass of Gimore et al..
}\label{table1}
\end{table}


The above relation is re-written as 
$$ R_1 = 6.5 \times 10^{-3} \times \left( \frac{M_{\rm mV}}{M_\odot} \right)^{0.55} 
\left(\frac{M/L}{M_\odot / L_\odot }\right)^{0.55} {\rm pc}
\eqno(4)
$$
 where $M_{\rm mV}$ is stellar mass and $M/L$ is mass-to-light ratio in the solar unit.
$M/L$ of a typical classical dSph is about 100. 
Its stellar mass measured from the absolute magnitude is about $10^6$ solar mass, 
and then the dark matter mass is about $10^8$ solar mass. 
If we adopt eq.(4), $R_1 \sim 100$ pc is predicted. Since radius of the typical classical dSph is on the order of 100 pc, 
we confirm the relation (4). 
In the last column of the table, we summarize $R_1$. 
Furthermore, regarding $r_{1/2}$ is not far from the realistic size of dSphs,
 we plot the ratio of $R_1/r_{1/2}$ in figure 1.
Interestingly, we find both of the radii are the same order.

To derive equation (4), we postulate that spatial distributions of stars and dark matter (DM) are almost the same. In the realistic situation, however, the spatial distribution of DM is wider than that of stars. Then, the DM gravitational potential of a realistic dSph is shallower than that of my model dSph when realistic and model dSphs resemble each other in stellar spatial extent and mass. At the same time, even though the spatial extents of model and realistic stellar distributions are alike, the shallowness of DM potential indicates a trend that stars in the realistic DM potential are distributed a bit more widely than those in the model DM potential. Thus $r_{1/2} > R_1$ is expected. 
Interestingly, $r_{1/2}$ is longer than $R_1$ except LeoII as presented in figure 1. The author hopes that this speculation on the overall trend of $r_{1/2} > R_1$ is checked by numerical simulations, and our simple approach is confirmed and revised.


Here, we estimate $M/L$ of SDSS J1257+3419, assuming
this is a dwarf spheroidal. The
absolute magnitude of the system is $-4.8$ in V-band. Then
we estimate the stellar mass of SDSS J1257+3419 to be 
$0.71 \times 10^4$ solar mass, assuming $M/L$ is unity in the solar unit for simplicity.
Although $r_{1/2}$ of SDSS J1257+3419 is observationally uncertain, 
 we think that the order of the observational value is reliable, and adopt $R_1 \sim 0.038$ kpc. 
Thus, we find its $M/L$ as about 994 which is expected $M/L$ from the Mateo plot. 
Our estimate suggests if SDSS J1257+3419 forms like old spheroids and becomes a dSph after the galactic wind epoch, 
its $M/L$ can reach at the order of 1000. 

A recent study (Simon \& Geha 2007) suggests $M/L$ of SDSS J1257+3419 is
$336 \pm 240$. Although its uncertainty is large because of long distance from the Milky Way, it is smaller than our estimate. We think the difference originates from the effect of the gradual mass-loss. The gradual mass-loss lessens absolute value of the initial binding energy $E_{\rm b}$ as long as the final radius is fixed (Hills 1980). That is, more realistic $M_{\rm DM}$ is smaller than the current estimate.
Unfortunately, numerical approach is necessary to quantitative estimate of the effect of the gradual mass-loss. This will be a next project.

By the way, SDSS J1257+3419 is very less massive. It may be disturbed 
by the tidal force from the Milky Way. 
Then,  
since the effect of dark matter is not considered in my previous paper (Kamaya 2007),
we re-check the possibility by the simple order of magnitude estimation.
Adopting a simple formula to estimate it, the tidal radius of SDSS J1257+3419 becomes $\sim$1000 pc when the Milky Way mass is about $10^{12}$ solar mass. 
Thus, its $r_{1/2}$ is found to be much shorter 
than the tidal radius. SDSS J1257+3419 can be self-gravitational system.
(e.g. Gonz$\rm \acute{a}$lez-Garc$\rm \acute{i}$a, Aguerri, \& Balcells 2005).
Thus, we can regard SDSS J1257+3419 as one of dSphs again.

The tidal effect can be important for the formation of a dSph. 
Mayer et al. (2001a,b) state in their paper 
that the strong tidal field of the Milky Way determines severe mass loss 
in their halos and disks and induces bar and bending instabilities 
that transform low surface-brightness dwarfs into dSphs. 
If this is true, 
SDSS J1257+3419 can also be a remnant of a low surface-brightness dwarf suffering the tidal stripping. 
Fortunately, once we remember that the tidal stripping enhances the mass loss effect owing to the galactic wind, 
 we find that this possibility is compatible with our conclusion.

\section{Summary}
The recent analysis of SDSS J1257+3419 has suggested that this stellar system
is either a faint and small dwarf galaxy or a faint and widely extended
globular cluster. The former possibility has been indicated because its mass density is similar to that of classical dSphs (kamaya 2007). Furthermore, 
according to the current estimate, $M/L$ of SDSS J1257+3419 is about 1000 which is expected for very small dSphs. 
As a result, the author insists this system is one of dSphs in the Milky Way system.


\end{document}